\documentstyle[10pt,epsf,epsfig,dp_delphititle,oldlfont]{dp_delphi}
\makeindex
\def\DpPaperGroup{EP}
\def\DpPaperRef{2001-004}
\def\DpDate{22-December-2000}
\def\DpAuthors{DELPHI Collaboration}
\def\DpSubmit{(Accepted by Phys.Lett.B)}
\def\DpTitle{{\bf Search for the Standard Model Higgs boson\\
                  at LEP in the year 2000}}
\def\DpComment{}
\def\DpEMail{ }

\begin{document}
\makeatletter
\newcount\@tempcntc
\def\@citex[#1]#2{\if@filesw\immediate\write\@auxout{\string\citation{#2}}\fi
  \@tempcnta\z@\@tempcntb\m@ne\def\@citea{}\@cite{\@for\@citeb:=#2\do
    {\@ifundefined
       {b@\@citeb}{\@citeo\@tempcntb\m@ne\@citea\def\@citea{,}{\bf ?}\@warning
       {Citation `\@citeb' on page \thepage \space undefined}}%
    {\setbox\z@\hbox{\global\@tempcntc0\csname b@\@citeb\endcsname\relax}%
     \ifnum\@tempcntc=\z@ \@citeo\@tempcntb\m@ne
       \@citea\def\@citea{,}\hbox{\csname b@\@citeb\endcsname}%
     \else
      \advance\@tempcntb\@ne
      \ifnum\@tempcntb=\@tempcntc
      \else\advance\@tempcntb\m@ne\@citeo
      \@tempcnta\@tempcntc\@tempcntb\@tempcntc\fi\fi}}\@citeo}{#1}}
\def\@citeo{\ifnum\@tempcnta>\@tempcntb\else\@citea\def\@citea{,}%
  \ifnum\@tempcnta=\@tempcntb\the\@tempcnta\else
   {\advance\@tempcnta\@ne\ifnum\@tempcnta=\@tempcntb \else \def\@citea{--}\fi
    \advance\@tempcnta\m@ne\the\@tempcnta\@citea\the\@tempcntb}\fi\fi}
 
\makeatother
\begin{titlepage}
\pagenumbering{roman}
\CERNpreprint{\DpPaperGroup}{\DpPaperRef} 
\date{{\small\DpDate}} 
\title{\DpTitle} 
\address{\DpAuthors} 
\begin{shortabs} 
\noindent
%
\noindent
Searches for the Standard Model Higgs boson
have been performed in the data collected by the DELPHI experiment
at LEP in the year 2000 at centre-of-mass energies 
between 200 and 209~\mbox{${\mathrm{GeV}} $}
corresponding to a total integrated \mbox{luminosity} of 
224~\mbox{pb$^{-1}$}.
No evidence for a Higgs signal is observed in the
kinematically accessible mass range,
and a 95\% CL lower mass limit of 114.3~\mbox{${\mathrm{GeV}}/c^2 $} is set,
to be compared with an expected median limit 
of 113.5~\mbox{${\mathrm{GeV}}/c^2 $} for 
these data.  
\end{shortabs}
\vfill
\begin{center}
\DpSubmit \ \\ 
\DpComment \ \\
\DpEMail \ \\
\end{center}
\vfill
\clearpage
\headsep 10.0pt
\addtolength{\textheight}{10mm}
\addtolength{\footskip}{-5mm}
\begingroup
%
\newcommand{\DpName}[2]{\hbox{#1$^{\ref{#2}}$},\hfill}
\newcommand{\DpNameTwo}[3]{\hbox{#1$^{\ref{#2},\ref{#3}}$},\hfill}
\newcommand{\DpNameThree}[4]{\hbox{#1$^{\ref{#2},\ref{#3},\ref{#4}}$},\hfill}
\newskip\Bigfill \Bigfill = 0pt plus 1000fill
\newcommand{\DpNameLast}[2]{\hbox{#1$^{\ref{#2}}$}\hspace{\Bigfill}}
%
\footnotesize
\noindent
\DpName{P.Abreu}{LIP}
\DpName{W.Adam}{VIENNA}
\DpName{T.Adye}{RAL}
\DpName{P.Adzic}{DEMOKRITOS}
\DpName{Z.Albrecht}{KARLSRUHE}
\DpName{T.Alderweireld}{AIM}
\DpName{G.D.Alekseev}{JINR}
\DpName{R.Alemany}{CERN}
\DpName{T.Allmendinger}{KARLSRUHE}
\DpName{P.P.Allport}{LIVERPOOL}
% ...
\normalsize
\endgroup
\titlefoot{Department of Physics and Astronomy, Iowa State
     University, Ames IA 50011-3160, USA
    \label{AMES}}
\titlefoot{Physics Department, Univ. Instelling Antwerpen,
     Universiteitsplein 1, B-2610 Antwerpen, Belgium \\
     \indent~~and IIHE, ULB-VUB,
     Pleinlaan 2, B-1050 Brussels, Belgium \\
     \indent~~and Facult\'e des Sciences,
     Univ. de l'Etat Mons, Av. Maistriau 19, B-7000 Mons, Belgium
    \label{AIM}}
\titlefoot{Physics Laboratory, University of Athens, Solonos Str.
     104, GR-10680 Athens, Greece
    \label{ATHENS}}
\titlefoot{Department of Physics, University of Bergen,
     All\'egaten 55, NO-5007 Bergen, Norway
    \label{BERGEN}}
\titlefoot{Dipartimento di Fisica, Universit\`a di Bologna and INFN,
     Via Irnerio 46, IT-40126 Bologna, Italy
    \label{BOLOGNA}}
\titlefoot{Centro Brasileiro de Pesquisas F\'{\i}sicas, rua Xavier Sigaud 150,
     BR-22290 Rio de Janeiro, Brazil \\
     \indent~~and Depto. de F\'{\i}sica, Pont. Univ. Cat\'olica,
     C.P. 38071 BR-22453 Rio de Janeiro, Brazil \\
     \indent~~and Inst. de F\'{\i}sica, Univ. Estadual do Rio de Janeiro,
     rua S\~{a}o Francisco Xavier 524, Rio de Janeiro, Brazil
    \label{BRASIL}}
\titlefoot{Comenius University, Faculty of Mathematics and Physics,
     Mlynska Dolina, SK-84215 Bratislava, Slovakia
    \label{BRATISLAVA}}
\titlefoot{Coll\`ege de France, Lab. de Physique Corpusculaire, IN2P3-CNRS,
     FR-75231 Paris Cedex 05, France
    \label{CDF}}
\titlefoot{CERN, CH-1211 Geneva 23, Switzerland
    \label{CERN}}
\titlefoot{Institut de Recherches Subatomiques, IN2P3 - CNRS/ULP - BP20,
     FR-67037 Strasbourg Cedex, France
    \label{CRN}}
\titlefoot{Now at DESY-Zeuthen, Platanenallee 6, D-15735 Zeuthen, Germany
    \label{DESY}}
\titlefoot{Institute of Nuclear Physics, N.C.S.R. Demokritos,
     P.O. Box 60228, GR-15310 Athens, Greece
    \label{DEMOKRITOS}}
\titlefoot{FZU, Inst. of Phys. of the C.A.S. High Energy Physics Division,
     Na Slovance 2, CZ-180 40, Praha 8, Czech Republic
    \label{FZU}}
\titlefoot{Currently at DPNC,
     University of Geneva,
     Quai Ernest-Ansermet 24, CH-1211, Geneva, Switzerland
    \label{GENEVA}}
\titlefoot{Dipartimento di Fisica, Universit\`a di Genova and INFN,
     Via Dodecaneso 33, IT-16146 Genova, Italy
    \label{GENOVA}}
\titlefoot{Institut des Sciences Nucl\'eaires, IN2P3-CNRS, Universit\'e
     de Grenoble 1, FR-38026 Grenoble Cedex, France
    \label{GRENOBLE}}
\titlefoot{Helsinki Institute of Physics, HIP,
     P.O. Box 9, FI-00014 Helsinki, Finland
    \label{HELSINKI}}
\titlefoot{Joint Institute for Nuclear Research, Dubna, Head Post
     Office, P.O. Box 79, RU-101 000 Moscow, Russian Federation
    \label{JINR}}
\titlefoot{Institut f\"ur Experimentelle Kernphysik,
     Universit\"at Karlsruhe, Postfach 6980, DE-76128 Karlsruhe,
     Germany
    \label{KARLSRUHE}}
\titlefoot{Institute of Nuclear Physics and University of Mining and Metalurgy,
     Ul. Kawiory 26a, PL-30055 Krakow, Poland
    \label{KRAKOW}}
\titlefoot{Universit\'e de Paris-Sud, Lab. de l'Acc\'el\'erateur
     Lin\'eaire, IN2P3-CNRS, B\^{a}t. 200, FR-91405 Orsay Cedex, France
    \label{LAL}}
\titlefoot{LIP, IST, FCUL - Av. Elias Garcia, 14-$1^{o}$,
     PT-1000 Lisboa Codex, Portugal
    \label{LIP}}
\titlefoot{Department of Physics, University of Liverpool, P.O.
     Box 147, Liverpool L69 3BX, UK
    \label{LIVERPOOL}}
\titlefoot{LPNHE, IN2P3-CNRS, Univ.~Paris VI et VII, Tour 33 (RdC),
     4 place Jussieu, FR-75252 Paris Cedex 05, France
    \label{LPNHE}}
\titlefoot{Department of Physics, University of Lund,
     S\"olvegatan 14, SE-223 63 Lund, Sweden
    \label{LUND}}
\titlefoot{Universit\'e Claude Bernard de Lyon, IPNL, IN2P3-CNRS,
     FR-69622 Villeurbanne Cedex, France
    \label{LYON}}
\titlefoot{Univ. d'Aix - Marseille II - CPP, IN2P3-CNRS,
     FR-13288 Marseille Cedex 09, France
    \label{MARSEILLE}}
\titlefoot{Dipartimento di Fisica, Universit\`a di Milano and INFN-MILANO,
     Via Celoria 16, IT-20133 Milan, Italy
    \label{MILANO}}
\titlefoot{Dipartimento di Fisica, Univ. di Milano-Bicocca and
     INFN-MILANO, Piazza delle Scienze 2, IT-20126 Milan, Italy
    \label{MILANO2}}
\titlefoot{IPNP of MFF, Charles Univ., Areal MFF,
     V Holesovickach 2, CZ-180 00, Praha 8, Czech Republic
    \label{NC}}
\titlefoot{NIKHEF, Postbus 41882, NL-1009 DB
     Amsterdam, The Netherlands
    \label{NIKHEF}}
\titlefoot{National Technical University, Physics Department,
     Zografou Campus, GR-15773 Athens, Greece
    \label{NTU-ATHENS}}
\titlefoot{Physics Department, University of Oslo, Blindern,
     NO-1000 Oslo 3, Norway
    \label{OSLO}}
\titlefoot{Dpto. Fisica, Univ. Oviedo, Avda. Calvo Sotelo
     s/n, ES-33007 Oviedo, Spain
    \label{OVIEDO}}
\titlefoot{Department of Physics, University of Oxford,
     Keble Road, Oxford OX1 3RH, UK
    \label{OXFORD}}
\titlefoot{Dipartimento di Fisica, Universit\`a di Padova and
     INFN, Via Marzolo 8, IT-35131 Padua, Italy
    \label{PADOVA}}
\titlefoot{Rutherford Appleton Laboratory, Chilton, Didcot
     OX11 OQX, UK
    \label{RAL}}
\titlefoot{Dipartimento di Fisica, Universit\`a di Roma II and
     INFN, Tor Vergata, IT-00173 Rome, Italy
    \label{ROMA2}}
\titlefoot{Dipartimento di Fisica, Universit\`a di Roma III and
     INFN, Via della Vasca Navale 84, IT-00146 Rome, Italy
    \label{ROMA3}}
\titlefoot{DAPNIA/Service de Physique des Particules,
     CEA-Saclay, FR-91191 Gif-sur-Yvette Cedex, France
    \label{SACLAY}}
\titlefoot{Instituto de Fisica de Cantabria (CSIC-UC), Avda.
     los Castros s/n, ES-39006 Santander, Spain
    \label{SANTANDER}}
\titlefoot{Dipartimento di Fisica, Universit\`a degli Studi di Roma
     La Sapienza, Piazzale Aldo Moro 2, IT-00185 Rome, Italy
    \label{SAPIENZA}}
\titlefoot{Inst. for High Energy Physics, Serpukov
     P.O. Box 35, Protvino, (Moscow Region), Russian Federation
    \label{SERPUKHOV}}
\titlefoot{J. Stefan Institute, Jamova 39, SI-1000 Ljubljana, Slovenia
     and Laboratory for Astroparticle Physics,\\
     \indent~~Nova Gorica Polytechnic, Kostanjeviska 16a, SI-5000 Nova Gorica, Slovenia, \\
     \indent~~and Department of Physics, University of Ljubljana,
     SI-1000 Ljubljana, Slovenia
    \label{SLOVENIJA}}
\titlefoot{Fysikum, Stockholm University,
     Box 6730, SE-113 85 Stockholm, Sweden
    \label{STOCKHOLM}}
\titlefoot{Dipartimento di Fisica Sperimentale, Universit\`a di
     Torino and INFN, Via P. Giuria 1, IT-10125 Turin, Italy
    \label{TORINO}}
\titlefoot{Dipartimento di Fisica, Universit\`a di Trieste and
     INFN, Via A. Valerio 2, IT-34127 Trieste, Italy \\
     \indent~~and Istituto di Fisica, Universit\`a di Udine,
     IT-33100 Udine, Italy
    \label{TU}}
\titlefoot{Univ. Federal do Rio de Janeiro, C.P. 68528
     Cidade Univ., Ilha do Fund\~ao
     BR-21945-970 Rio de Janeiro, Brazil
    \label{UFRJ}}
\titlefoot{Department of Radiation Sciences, University of
     Uppsala, P.O. Box 535, SE-751 21 Uppsala, Sweden
    \label{UPPSALA}}
\titlefoot{IFIC, Valencia-CSIC, and D.F.A.M.N., U. de Valencia,
     Avda. Dr. Moliner 50, ES-46100 Burjassot (Valencia), Spain
    \label{VALENCIA}}
\titlefoot{Institut f\"ur Hochenergiephysik, \"Osterr. Akad.
     d. Wissensch., Nikolsdorfergasse 18, AT-1050 Vienna, Austria
    \label{VIENNA}}
\titlefoot{Inst. Nuclear Studies and University of Warsaw, Ul.
     Hoza 69, PL-00681 Warsaw, Poland
    \label{WARSZAWA}}
\titlefoot{Fachbereich Physik, University of Wuppertal, Postfach
     100 127, DE-42097 Wuppertal, Germany
    \label{WUPPERTAL}}
\addtolength{\textheight}{-10mm}
\addtolength{\footskip}{5mm}
\clearpage
\headsep 30.0pt
\end{titlepage}
%
\pagenumbering{arabic} 
\setcounter{footnote}{0} %
\large
%
\newcommand{\Zz} {\mbox{Z}}
\newcommand{\Zn} {\mbox{$ {\mathrm Z}^0 \,$}}
\newcommand{\Wp} {\mbox{$ {\mathrm W}^+ \,$}}
\newcommand{\W} {\mbox{$ {\mathrm W}^{\pm} \,$}}
\newcommand{\Hz} {\mbox{H}}
\newcommand{\hz} {\mbox{h}}
\newcommand{\hp} {\mbox{$ {\mathrm H}^+ \,$}}
\newcommand{\hm} {\mbox{$ {\mathrm H}^- $}}
\newcommand{\hpm}{\mbox{${\mathrm H}^{\pm}$}}
\newcommand{\tol}{\mbox{$\tau$ }}
\newcommand{\MZ} {\mbox{$ m_{\mathrm Z} \, $}}
\newcommand{\MW} {\mbox{$ m_{\mathrm W} \, $}}
\newcommand{\MA} {\mbox{$ m_{\mathrm A} \, $}}
\newcommand{\MAtbeta} {$ m_{\mathrm A}/\tan \beta$}
\newcommand{\MH} {\mbox{$ m_{\mathrm H} \, $}}
\newcommand{\MT} {\mbox{$ m_{\mathrm t} $}}
\newcommand{\mh} {\mbox{$ m_{\mathrm h} $}}
\newcommand{\mH} {\mbox{$ m_{{\mathrm H}^{\pm}}$}}
\newcommand{\ee}{\mbox{${\mathrm e}^+{\mathrm e}^-$}}
\newcommand{\mm}{\mbox{$\mu^+ \mu^-$}}
\newcommand{\ffbar}{\mbox{${\mathrm f}\bar{\mathrm f}$} }
\newcommand{\qqbar}{\mbox{${\mathrm q}\bar{\mathrm q}$} }
\newcommand{\bbbar}{\mbox{${\mathrm b}\bar{\mathrm b}$} }
\newcommand{\ccbar}{\mbox{${\mathrm c}\bar{\mathrm c}$} }
\newcommand{\nunubar}{$\nu \bar{\nu}\;$}
\newcommand{\ton}{$\tau \nu_{\tau} \;$ }
\newcommand{\toto}{\mbox{$\tau^+ \tau^-$}}
\newcommand{\aju}{\alpha_{1}^{jet}}
\newcommand{\ajd}{\alpha_{2}^{jet}}
\newcommand{\Mmm}{$M_{\mu \mu}$ }
\newcommand{\Mrec}{$M_{rec}$ }
\newcommand{\doubl}{$\Gamma_{b\bar{b}}/\Gamma_{had}$}
\newcommand{\btagpe}{\mbox{$P_{\mathrm E}$} }
\newcommand{\btagpep}{\mbox{$P^{+}_{\mathrm E}$} }
\newcommand{\fthvis}{\mbox{$\theta^f_{vis}$}}
\newcommand{\xb}{\mbox{$x_{\mathrm b}$}}
\newcommand{\xbi}{\mbox{$x_{\mathrm b}^{i}$}}
\newcommand{\hmm}{\mbox{\Hz$ \mu^+ \mu^-$}}
\newcommand{\hee}{\mbox{\Hz${\mathrm {e^+ e^-}}$}}
\newcommand{\hnn}{\mbox{\Hz$ \nu \bar{\nu}$}}
\newcommand{\hqq}{$\Hz{\mathrm {q \bar{q}}}$ }
\newcommand{\htt}{$({\mathrm \Hz \rightarrow q \bar{q}})\tau^+\tau^-$ }
\newcommand{\ttZ}{$({\Hz }\rightarrow\tau^+\tau^-) {\mathrm q \bar{q}}$ }
\newcommand{\ttqq}{$\tau^+\tau^- {\mathrm q \bar{q}}$ }
\newcommand{\hAtt}{${\mathrm {hA}} \rightarrow \tau^+\tau^- {\mathrm q \bar{q}}$ }
\newcommand{\hAbb}{hA$\rightarrow $\bbbar \bbbar }
\newcommand{\hAA} {${\mathrm{h}}\rightarrow {\mathrm{AA}}$}
\newcommand{\bbg} {${\mathrm b \bar{b}}(\gamma)$ }
\newcommand{\qqg} {\mbox{$ {\mathrm q}\bar{\mathrm q}(\gamma) $}}
\newcommand{\qqgg}{${\mathrm q \bar{q}} gg\;$}
\newcommand{\gaga}{\mbox{$\gamma \gamma$ }}
\newcommand{\gghad}{$\gamma \gamma \rightarrow {\rm hadrons}$ }
\newcommand{\llbar}{\mbox{${\mathrm l^+ l^- }(\gamma)$ }}
\newcommand{\llg} {\mbox{${\mathrm ll}(\gamma)$}}
\newcommand{\eeg} {\mbox{${\mathrm e^+ e^- }(\gamma)$ }}
\newcommand{\WW} {\mbox{${\mathrm W}^+{\mathrm W}^-$}}
\newcommand{\WWb} {${\mathrm {WW}}$ }
\newcommand{\Wen} {${\mathrm {We}\nu}$ }
\newcommand{\Zee} {${\mathrm {Zee}}$ }
\newcommand{\ZZ} {${\mathrm {ZZ}}$}
\newcommand{\ZH} {\mbox{${\mathrm {HZ}}$}}
\newcommand{\hA} {\mbox{$ {\mathrm h} {\mathrm A} \,$}}
\newcommand{\hZ} {\mbox{$ {\mathrm h} {\mathrm Z} \,$}}
\newcommand{\eeww} {\mbox{\ee $\rightarrow$ \WW}}
\newcommand{\eezz} {\mbox{\ee $\rightarrow {\mathrm ZZ}$}}
\newcommand{\eezee} {\mbox{\ee $\rightarrow $ \Zee}}
\newcommand{\eewenu} {\mbox{\ee $\rightarrow $ \Wen}}
\newcommand{\eehz} {\mbox{\ee $\rightarrow $ \hZ }}
\newcommand{\eehA} {\mbox{\ee $\rightarrow $ \hA }}
\newcommand{\eehpm} {\mbox{\ee $\rightarrow $\ HH }}
\newcommand{\eeqq} {\mbox{\ee \qqbar }}
\newcommand{\mmqq} {\mbox{$\mu^+ \mu^- $\qqbar }}
\newcommand{\tautauqq}{\mbox{$\tau^+ \tau^- $\qqbar }}
\newcommand{\evqq} {\mbox{${\mathrm e} \nu $\qqbar }}
\newcommand{\llqq} {\mbox{${\ell^+\ell^- }$\qqbar }}
\newcommand{\lvqq} {\mbox{${\mathrm l^+} \nu$\qqbar }}
\newcommand{\tauvqq} {\mbox{$\tau\nu$${\mathrm q'}\bar{\mathrm q}$ }}
\newcommand{\qqqq} {\mbox{\qqbar\qqbar}}
\newcommand{\tbeta} {\mbox{$\tan \beta$}}
\newcommand{\MeV} {\mbox{${\mathrm{MeV}} $}}
\newcommand{\MeVc} {\mbox{${\mathrm{MeV}}/c $}}
\newcommand{\MeVcc} {\mbox{${\mathrm{MeV}}/c^2 $}}
\newcommand{\GeV} {\mbox{${\mathrm{GeV}} $}}
\newcommand{\GeVc} {\mbox{${\mathrm{GeV}}/c $}}
\newcommand{\GeVcc} {\mbox{${\mathrm{GeV}}/c^2 $}}
\newcommand{\TeVcc} {\mbox{${\mathrm{TeV}}/c^2 $}}
\newcommand{\dgree} {\mbox{$^{\circ}$}}
\newcommand{\mydeg} {$^{\circ}$}
\newcommand{\Zvv } {\mbox{$ Z \nu \bar{\nu} $}}
\newcommand{\pbinv} {\mbox{pb$^{-1}$}}
\newcommand{\mhp }{\mbox{$ m_{{\mathrm H}^+} \, $}}
\newcommand{\HH }{\mbox{$ {\mathrm H}^+{\mathrm H}^- $}}
\newcommand{\hptn }{\mbox{$ \hp \rightarrow \tau^+ \nu_{\tau} \,$}}
\newcommand{\hpcs }{\mbox{$ \hp \rightarrow c \bar{s} \,$}}
\newcommand{\hpcb }{\mbox{$ \hp \rightarrow c \bar{b} \,$}}
\newcommand{\cscs }{\mbox{$c \bar{s} \bar{c} s \,$}}
\newcommand{\tntn }{\mbox{$\tau^+ \nu_{\tau} \tau^- {\bar{\nu}}_{\tau} \,$}}
\newcommand{\cstn }{\mbox{$c s \tau \nu_{\tau} \,$}}
\newcommand{\hpmtntn }{\mbox{$ \HH \rightarrow \tntn \,$}}
\newcommand{\hpmcstn }{\mbox{$ \HH \rightarrow \cstn \,$}}
\newcommand{\hpmcscs }{\mbox{$ \HH \rightarrow \cscs \,$}}
\newcommand{\mhrec }{\mbox{$ m_{{\mathrm H}^+}^{rec} \,$}}
\newcommand{\fcstn }{\mbox{$ F_{cs\tau\nu}\,$}}
\newcommand{\fcscs }{\mbox{$ F_{cscs}\,$}}
\newcommand{\fthsph }{\mbox{$ \theta^f_{sph} \,$}}
\newcommand{\thetast }{\mbox{$ \theta^* \,$}}
\newcommand{\bmplane }{\mbox{$\mhp,{\mathrm Br (H^+ \rightarrow leptons)}$}}
\newcommand{\sqrts }{\mbox{$ \sqrt{s} \,$}}
\newcommand{\like}{\mbox{$\cal L$}}
\newcommand{\likear}{\mbox{$\cal Q$}}
\newcommand{\emin}{\mbox{$E_{\mathrm min}$}}
\newcommand{\emax}{\mbox{$E_{\mathrm max}$}}
\newcommand{\alphamin}{\mbox{$\alpha_{\mathrm min}$}}
\newcommand{\betamin}{\mbox{$\beta_{\mathrm min}$}}
\newcommand{\clsinf}{\mbox{$\langle CL_s \rangle$}}
\newcommand{\GF} {\mbox{$ {\mathrm G}_{\mathrm F} $}}
\newcommand{\GZ} {\mbox{$ \Gamma_{{\mathrm Z}^0} $}}
\newcommand{\GW} {\mbox{$ \Gamma_{\mathrm W} $}}
\newcommand{\ro} {\mbox{$ \frac{m^{2}_{{\mathrm W}}}{m^{2}_{{\mathrm Z}}
\,\cos^{2}\theta_{\mathrm W}}\, $}}
\newcommand{\sw} {\mbox{$ \sin\theta_{\mathrm W} $}}
\newcommand{\ssw} {\mbox{$ \sin^{2} \theta_{\mathrm W} $}}
\newcommand{\cw} {\mbox{$ \cos\theta_{\mathrm W} $}}
\newcommand{\thw} {\mbox{$ \theta_{\mathrm W} $}}
\newcommand{\alphmz} {\mbox{$ \alpha (m_{\mathrm Z}) $}}
\newcommand{\alphas} {\mbox{$ \alpha_{\mathrm s} $}}
\newcommand{\alphmsb} {\mbox{$ \alphas (m_{\mathrm Z})
_{\overline{\mathrm{MS}}} $}}
\newcommand{\alphsbar} {\mbox{$ \overline{\alpha}_{\mathrm s} $}}
\newcommand{\HZ} {\mbox{$ {\mathrm H}^0 {\mathrm Z}^0 $}}
\newcommand{\mumu} {\mbox{$ \mu^+ \mu^- $}}
\newcommand{\ffb} {\mbox{$ {\mathrm f}\bar{{\mathrm f}}
({\mathrm n}\gamma) $}}
\newcommand{\zee} {\mbox{$ {{\mathrm Ze}}^+{{\mathrm e}}^- $}}
\newcommand{\ewn} {\mbox{$ {\mathrm{W e}} \nu_{\mathrm e} $}}
\newcommand{\qaqb} {\mbox{$ {\mathrm q}_1 \bar{\mathrm q}_2 $}}
\newcommand{\qcqd} {\mbox{$ {\mathrm q}_3 \bar{\mathrm q}_4 $}}
\newcommand{\eeffb} {\mbox{$ \ee \rightarrow \ffb \,$}}
\newcommand{\eeee} {\mbox{$ \ee \rightarrow \ee \,$}}
\newcommand{\eeggqpm} {\mbox{$ \ee \rightarrow \ggqpm \,$}}
\newcommand{\eeggvdm} {\mbox{$ \ee \rightarrow \ggvdm \,$}}
\newcommand{\eeggqcd} {\mbox{$ \ee \rightarrow \ggqcd \,$}}
\newcommand{\ggee} {\mbox{$ \gaga \rightarrow \ee \,$}}
\newcommand{\ggmm} {\mbox{$ \gaga \rightarrow \mumu \,$}}
\newcommand{\ggtt} {\mbox{$ \gaga \rightarrow \toto \,$}}
\newcommand{\ggh} {\mbox{$ \gaga \rightarrow {\mathrm hadrons} \,$}}
\newcommand{\Ecms} {\mbox{$ E_{\mathrm{cms}} \,$}}
\newcommand{\Evis} {\mbox{$ E_{\mathrm{vis}} \,$}}
\newcommand{\Etot} {\mbox{$ E_{\mathrm{tot}} \,$}}
\newcommand{\Ecal} {\mbox{$ E_{\mathrm{calo}} \,$}}
\newcommand{\Echa} {\mbox{$ E_{\mathrm{ch}} \,$}}
\newcommand{\Esh} {\mbox{$ E_{\mathrm{shower}} \,$}}
\newcommand{\Eforw} {\mbox{$ E_{\mathrm{forward}} \,$}}
\newcommand{\Mvis} {\mbox{$ M_{\mathrm{vis}} \,$}}
\newcommand{\pvis} {\mbox{$ p_{\mathrm{vis}} \,$}}
\newcommand{\Minv} {\mbox{$ M_{\mathrm{inv}} \,$}}
\newcommand{\ymin} {\mbox{$ y_{cut} \,$}}
\newcommand{\mcha} {\mbox{$ M_{\mathrm{ch}} \,$}}
\newcommand{\acol} {\mbox{$ \cal{A} \,$}}
\newcommand{\Ej} {\mbox{$ E_j \,$}}
\newcommand{\alfij} {\mbox{$ \alpha_{ij} \,$}}
\newcommand{\pt} {\mbox{$ P^{T}_{\mathrm{vis}} \,$}}
\newcommand{\tvis} {\mbox{$ \theta_{\mathrm{vis}} \,$}}
\newcommand{\actvis} {\mbox{$ |\cos(\tvis)| \,$}}
\newcommand{\tsph} {\mbox{$ \theta_{sph} \,$}}
\newcommand{\acosph} {\mbox{$ |\cos(\theta_{sph})| \,$}}
\newcommand{\khic} {\mbox{$ \chi^2 \,$}}
\newcommand{\mulcmin} {\mbox{$ min(M_{\mathrm{ch}}^{\mathrm{jet}}) \,$}}
\newcommand{\vniels} {\mbox{$ {\mathrm min}_{j}(E_j) \times
{\mathrm min}_{ij}(\alpha_{ij}) \,$}}
\newcommand{\Ptau} {\mbox{$ P_{\tau} $}}
\newcommand{\mean}[1] {\mbox{$ \left\langle #1 \right\rangle $}}
\newcommand{\phistar} {\mbox{$ \phi^* $}}
\newcommand{\thetapl} {\mbox{$ \theta_+ $}}
\newcommand{\phipl} {\mbox{$ \phi_+ $}}
\newcommand{\thetamin}{\mbox{$ \theta_- $}}
\newcommand{\phimin} {\mbox{$ \phi_- $}}
\newcommand{\ds} {\mbox{$ {\mathrm d} \sigma $}}
\newcommand{\jjlv} {\mbox{$ j j \ell \nu $}}
\newcommand{\jjjj} {\mbox{$ j j j j $}}
\newcommand{\jjvv} {\mbox{$ j j \nu \bar{\nu} $}}
\newcommand{\jjll} {\mbox{$ j j \ell \bar{\ell} $}}
\newcommand{\lvlv} {\mbox{$ \ell \nu \ell \nu $}}
\newcommand{\dz} {\mbox{$ \delta g_{\mathrm{W W Z} } $}}
\newcommand{\ptr} {\mbox{$ p_{\perp} $}}
\newcommand{\ptrjet} {\mbox{$ p_{\perp {\mathrm{jet}}} $}}
\newcommand{\gamgam} {\mbox{$ \gamma \gamma $}}
\newcommand{\djoin} {\mbox{$ d_{\mathrm{join}} $}}
\newcommand{\mErad} {\mbox{$ \left\langle E_{\mathrm{rad}} \right\rangle $}}
\newcommand{\Zto} {\mbox{$ {\mathrm Z} \to $}}
\def\NPB#1#2#3{{\it Nucl.~Phys.} {\bf{B#1}} (19#2) #3}
\def\PLB#1#2#3{{\it Phys.~Lett.} {\bf{B#1}} (19#2) #3}
\def\PRD#1#2#3{{\it Phys.~Rev.} {\bf{D#1}} (19#2) #3}
\def\PRL#1#2#3{{\it Phys.~Rev.~Lett.} {\bf{#1}} (19#2) #3}
\def\ZPC#1#2#3{{\it Z.~Phys.} {\bf C#1} (19#2) #3}
\def\PTP#1#2#3{{\it Prog.~Theor.~Phys.} {\bf#1} (19#2) #3}
\def\MPL#1#2#3{{\it Mod.~Phys.~Lett.} {\bf#1} (19#2) #3}
\def\PR#1#2#3{{\it Phys.~Rep.} {\bf#1} (19#2) #3}
\def\RMP#1#2#3{{\it Rev.~Mod.~Phys.} {\bf#1} (19#2) #3}
\def\HPA#1#2#3{{\it Helv.~Phys.~Acta} {\bf#1} (19#2) #3}
\def\NIMA#1#2#3{{\it Nucl.~Instr.~and~Meth.} {\bf#1} (19#2) #3}
\newcommand{\rs}{\mbox{$\sqrt{s}$}}
\newcommand{\hu}{\rule{0ex}{3ex}}
\newcommand{\AZ} {\mbox{$ {\mathrm A} {\mathrm Z} \, $}}
\newcommand{\hH} {\mbox{$ {\mathrm h} {\mathrm H} \, $}}

\section{Introduction}

The LEP accelerator was successfully operated at \ee\ collision 
energies up to 209~\GeV\ during the year 2000.
The DELPHI experiment has collected 
more than 224~\pbinv\ at centre-of-mass energies above 200~\GeV,
extending the range of searches for the Standard Model Higgs boson 
above the previous limits obtained by DELPHI~\cite{pap99,pap98,pap97},
by the other LEP collaborations, 
and by their combination by the LEP Higgs Working Group~\cite{adlo-cernep}.

The results shown in this letter are based on the detector calibration obtained
shortly after the end of data taking. 
They will be included in the preliminary combination of the 
LEP collaborations results on the 2000 year data~\cite{alo_2000},
being prepared by the LEP Higgs Working
group~\cite{LEPHWG_2000}.

\subsection{Data and simulation samples}

The data used in this analysis, corresponding to 
a total of 224.1~\pbinv\ collected by the DELPHI detector in 2000,
were analysed in the following subsamples: 
2.3~\pbinv\ at an average centre-of-mass energy of 202.6~\GeV, 
6.7~\pbinv\ at 203.9~\GeV, 
10.5~\pbinv\ at 204.8~\GeV,
62.5~\pbinv\ at 205.2~\GeV, 
18.2~\pbinv\ at 206.2~\GeV, 
115.2~\pbinv\ at 206.7~\GeV\
and 8.7~\pbinv\ at 208.2~\GeV.


Monte Carlo samples for background events were produced 
at fixed centre-of-mass
energies of 202, 204, 205, 206, 207 and 208~\GeV\
using the same simulation setup as for the 1999 analysis~\cite{pap99}. 
The samples correspond to about 200 times the collected luminosity. 

Similarly, signal events were produced using 
the {\tt HZHA}~\cite{hzha} generator,
varying the Higgs boson mass from 85~\GeVcc~ to
120~\GeVcc~ in 5~\GeVcc~ steps,
plus a fine scan in the most interesting zone, with 
samples simulated for mass hypotheses 108, 110, 112, 114 and 115~\GeVcc.

\subsection{Detector overview}

A detailed description of the DELPHI apparatus can be found in
\cite{DELPHI}. For the first three quarters of the year the detector was
operated in nominal conditions.

Data collected after the 1st of September, corresponding to the last
60~\pbinv, were affected by the complete failure of 
one sector (S6) of the TPC detector,
which amounts to 1/12 of the TPC acceptance.
Charged particle tracks crossing this sector were reconstructed using
the information from the Vertex, Inner and Outer detectors,
so the effect on the efficiency is limited.
A complete sample of background and signal channels simulated
with this TPC sector off was used 
to incorporate the small effect on the 
reconstructed event kinematics
and the impact on the ${\mathrm b}$-tagging efficiency
into the analysis of this data sample. 



To follow
more precisely the change of conditions during the data taking, 
the calibration of the impact parameter resolution
was performed with the high energy four-jet events. The same procedure
was applied to the simulation where the four-jet events were selected
with the same criteria and appropriately weighted according to
the predicted cross-sections of the corresponding processes.

The calibration of the ${\mathrm b}$-tagging used the tracks 
with negative impact
parameter, while only the tracks with positive impact parameter were
used in the lifetime based ${\mathrm b}$-tagging.

The number of tracks with negative impact parameter is not
affected by the calibration, and is used as further information
in the ${\mathrm b}$-tagging.
Therefore this calibration
procedure is not correlated with the physics measurement, while it 
improves significantly the agreement between data and simulation.

\begin{figure}[htbp]
\begin{center}
\epsfig{figure=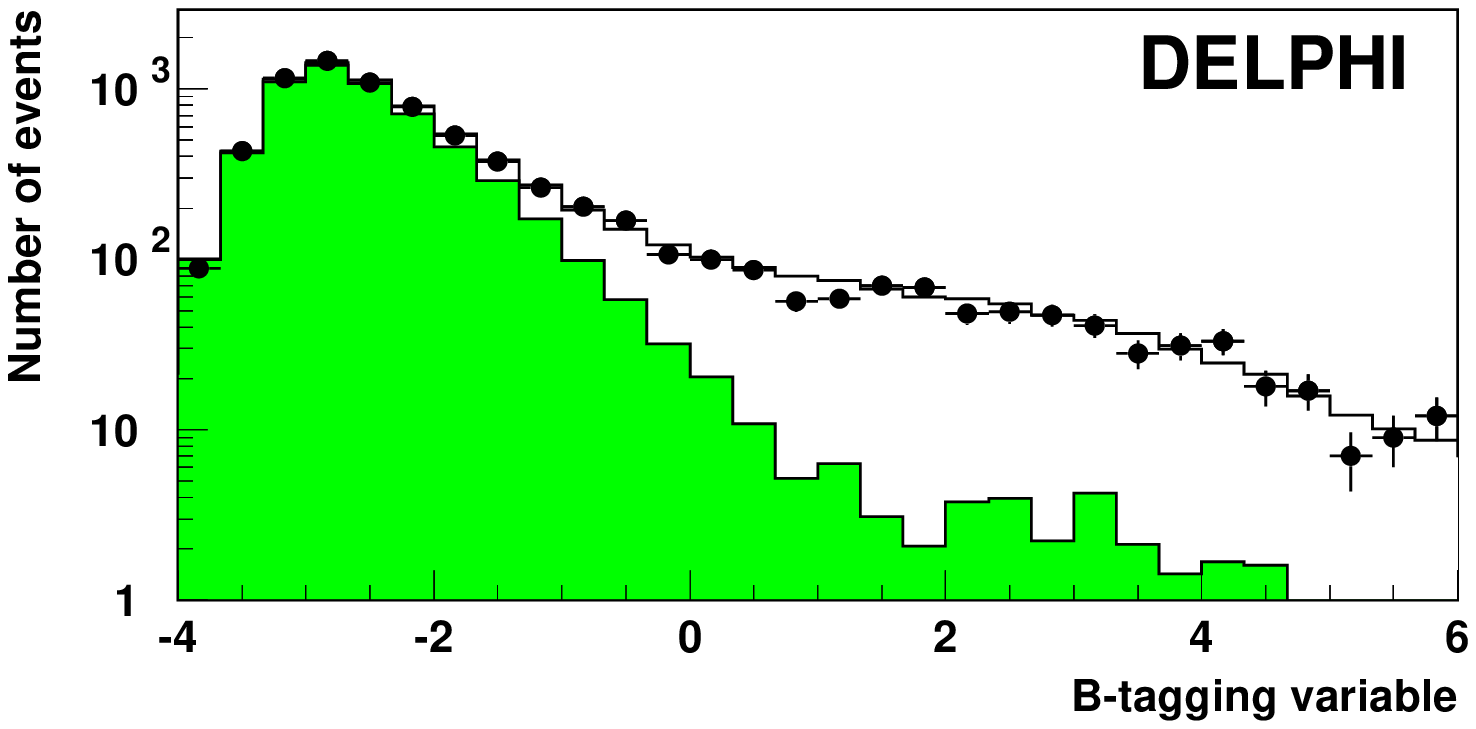,width=17cm} 
\epsfig{figure=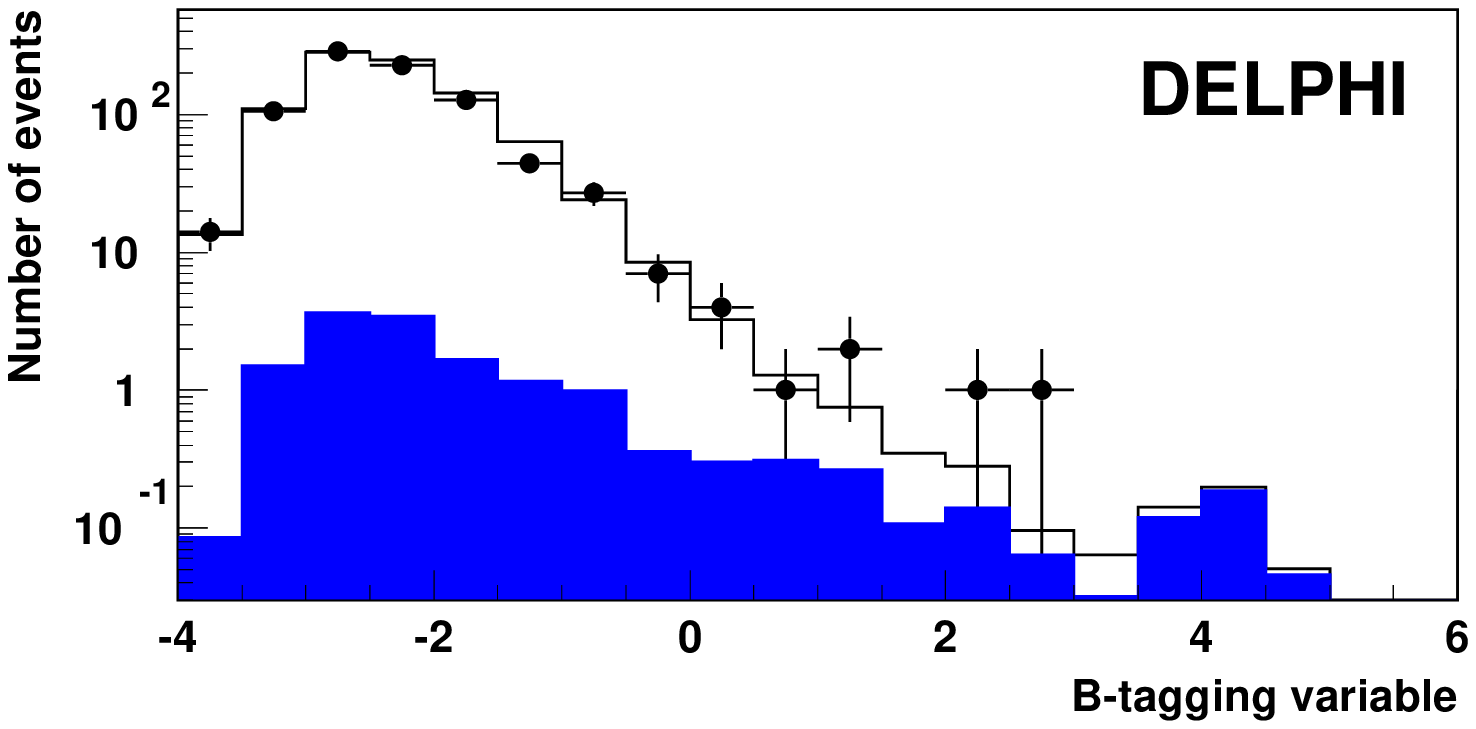,width=17cm} 
\caption[]{Top: distributions of the combined ${\mathrm b}$-tagging variable, 
for the year 2000 radiative return \Zz $\gamma$ data 
(dots) and simulation (histogram). The expected contribution 
of udsc-quarks and non-\qqbar $\gamma$ background is shown as the 
dark histogram. 

Bottom: same distribution for semileptonic \WW\ high 
energy events in the 2000 data. The shaded histogram corresponds to
the expected contribution from other processes, and shows 
the high purity of the selection.}
\label{fig:btagging}
\end{center}
\end{figure}

The overall performance of the combined ${\mathrm b}$-tagging
in  hadronic radiative return events (\mbox{\ee $\rightarrow$ \Zn $\gamma$}),
collected during the year 2000,
is illustrated in Fig.~\ref{fig:btagging}. 
Effects of possible imperfect modelling of the high b-tag tail 
from non-b quarks
were checked using the high energy semileptonic \WW\ data 
and are also shown in Fig.~\ref{fig:btagging}.

\section{Standard Model Higgs search}

The previous LEP combined limit\cite{adlo-cernep} on the Higgs mass at 95\% 
CL was close to 108~\GeVcc.  
Given the integrated luminosity corresponding to the 
data taken in the year 2000, 
the analysis is expected to cover efficiently the mass range
up to the kinematical limit allowed by the increase in 
centre-of-mass energy.


The following improvements for this high mass range have been introduced 
in the analysis of the two main channels.

The four-jet analysis benefits from a better tuned 
${\mathrm b}$-tagging procedure
and although it keeps the same event variables in the
analysis, the discriminant neural network has been optimized 
for the high mass hypotheses. 

The missing-energy channel includes 
a tighter preselection and additional variables in the likelihood,
resulting in a better background rejection for a high mass Higgs;
it is described in the following section.

\subsection{H\nunubar channel}

In this channel both the preselection and the final discriminating likelihood  
have been reoptimised in the spirit of a ``background free''
analysis. A set of stringent cuts~\cite{osaka00}
was applied prior to the construction of the likelihood.

 
The discriminating likelihood includes six variables defined 
after forcing the event into a two-jet configuration with the 
DURHAM~\cite{DURHAM} algorithm:
acoplanarity, acollinearity, 
polar angle of the missing momentum with respect to the beam direction, 
${\mathrm b}$-tagging, 
invariant mass in the transverse plane, 
the minimum of the energies 
around the most isolated particle and 
around the most energetic particle (normalised to their own energy). 
Three more variables are  defined  
leaving the number of jets 
free in the DURHAM algorithm with $y_{cut}=0.005$:
the minimum angle between the jet directions and the missing momentum 
in the transverse plane, the minimal jet charged multiplicity, and the 
maximum track or reconstructed lepton transverse momentum with respect to the jet axis.

The effect of the preselection on data and simulated samples is 
shown in Table~\ref{ta:hzsum205}. 
After a tighter cut to select the most significant candidates,
three candidate events remain, while 4.9 are expected according 
to the background simulation. 

\begin{figure}[htbp]
\begin{center}
\epsfig{figure=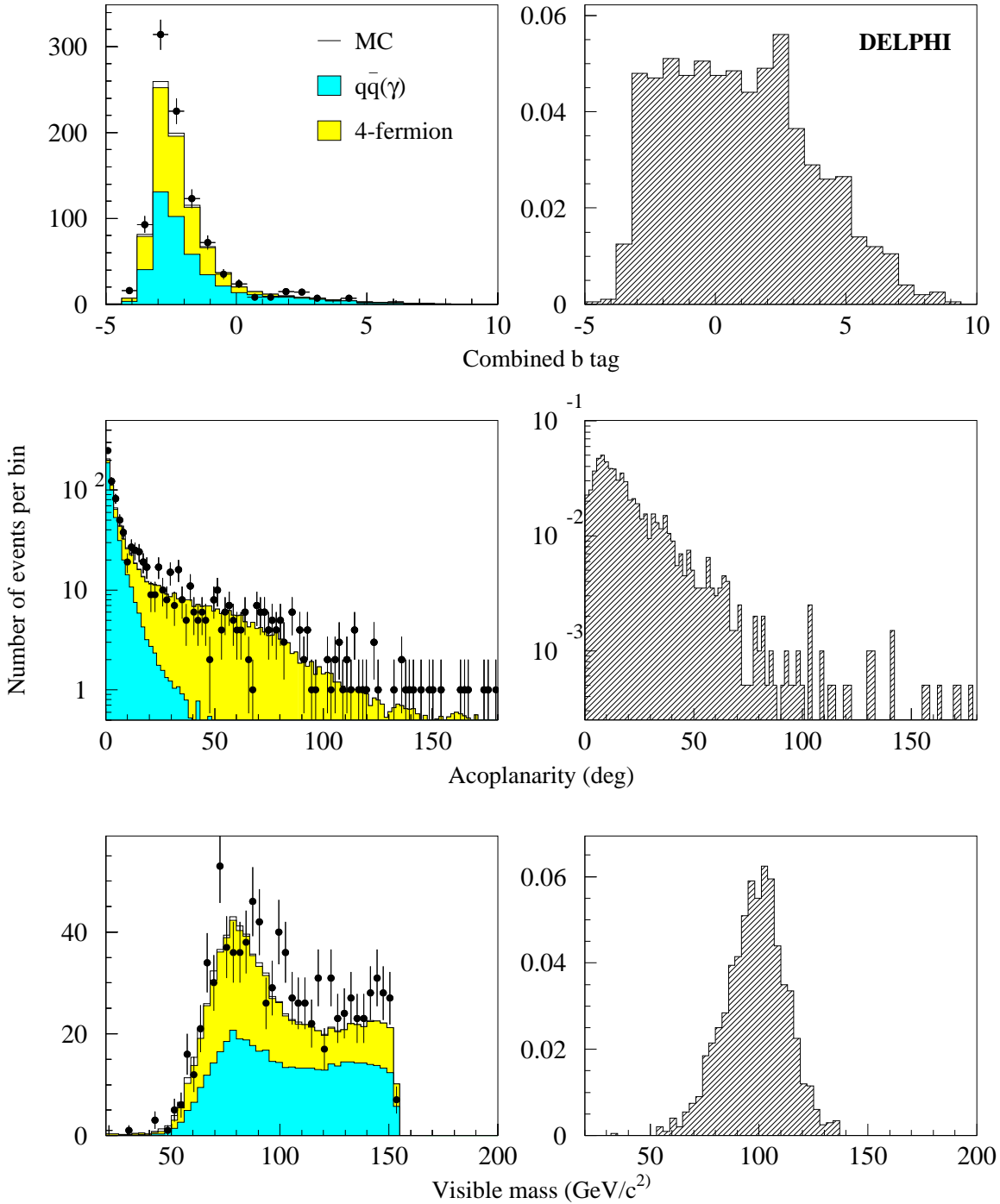,width=17cm}
\caption{ \hnn\ channel: 
  distributions of relevant analysis variables, at the preselection level. 
  Data at \rs~=~200-209~\GeV\ (dots) are compared with Standard Model
  background 
  expectations (left-hand side histograms) and with the expected distribution
  for a 114~\GeVcc\ Higgs mass signal (right-hand side histogram).}
\label{fig:hnunu_1}
\end{center}
\end{figure}

\begin{figure}[htbp]
\begin{center}
\epsfig{figure=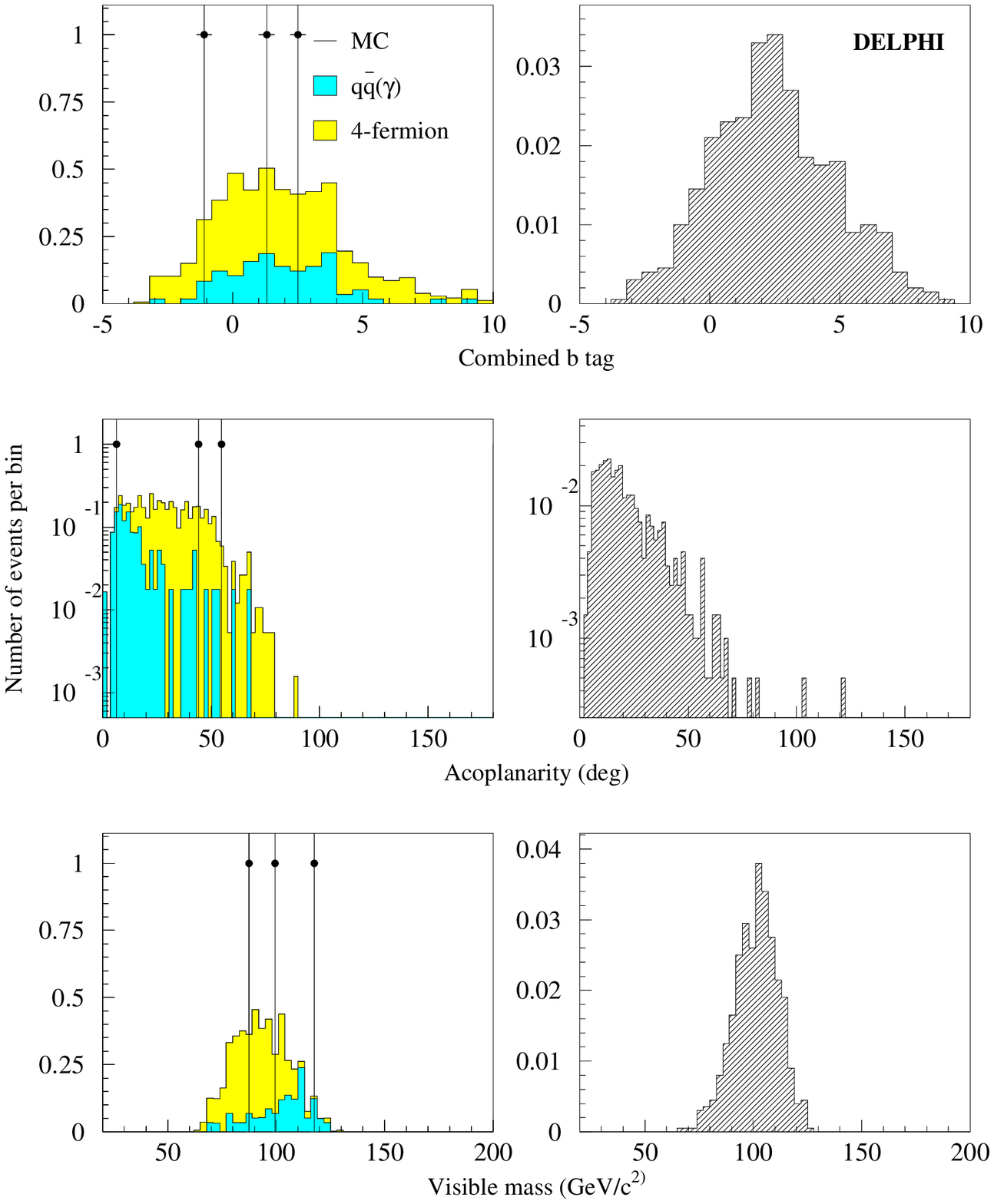,width=17cm}
\caption{ \hnn\ channel: 
same distributions as in Fig.\ref{fig:hnunu_1} but at the tight selection level.}
\label{fig:hnunu_2}
\end{center}
\end{figure}

Distributions for the most relevant variables in this analysis 
are shown both at preselection level (Fig.~\ref{fig:hnunu_1}), 
and at the tight selection level (Fig.~\ref{fig:hnunu_2}).

The reconstructed Higgs boson mass is defined as 
the visible mass given by a one-constraint fit where the recoil system 
is assumed to be an on-shell \Zn\ boson. It is used, 
together with the discriminant likelihood,
in the two-dimensional computation of the confidence levels 
for the Higgs hypotheses.

\vspace{2cm}

\begin{table}[hbtp]
\begin{center}
\begin{tabular}{cccccc}     \hline
Selection & Data & Background  & 
{\mbox{$ {\mathrm {\it q}}\bar{\mathrm {\it q}}(\gamma) $}}
& {\it 4-fermion} & Efficiency \\
\hline \hline
\multicolumn{6}{c} {\hnn\ channel} \\ \hline
preselection     &  970 &  880   &  {\it 467}   &  {\it 390}   &   67\% \\
candidates selection  &   90 &  99.7  &  {\it 50.4}  &   {\it 49.3}  &   60\% \\ 
tight selection  &    3 &   4.9  &  {\it  1.4}  &   {\it 3.5}   &   30\% \\ 
\hline
\multicolumn{6}{c} {\hee\ channel } \\ \hline
preselection     & 1242 &  1172   & {\it 745 }  & {\it 416  } &   78\% \\
candidates selection  &    7 &  11.6   & {\it  0.5}  & {\it  10.4} &   57\% \\ 
tight selection  &    1 &   3.5   & {\it  0.1}  & {\it   3.2} &   49\% \\ 
\hline
\multicolumn{6}{c} {\hmm\ channel} \\ \hline
preselection     & 3780 &  3763    & {\it 2671} & {\it 1067} &  81\% \\
candidates selection  &    7 &   10.6   & {\it  0.2} & {\it 10.4} &  67\% \\ 
tight selection  &    2 &    3.6   & {\it  0.1} & {\it  3.5} &  56\% \\ 
\hline
\multicolumn{6}{c} {\ttqq\ channel } \\ \hline
preselection     & 9180  & 8913    & {\it 5425} & {\it  3468}   &  98\% \\
candidates selection  &    5 &   6.0    & {\it  0.4} & {\it   5.6}   &  22\% \\
tight selection  &    2 &   4.1   &  {\it  0.1} & {\it   4.0}   &  19\% \\
\hline
\multicolumn{6}{c} {\hqq\ channel} \\ \hline
preselection     &  2266 &  2342      & {\it  680  }  & {\it 1662 }  &   85\% \\
candidates selection  &   398 &   423.7    & {\it  154.9}  & {\it 268.8} &   79\% \\ 
tight selection  &     8 &     7.4    & {\it    2.8}  & {\it   4.6} &   36\% \\ 
\hline
\vspace{0.5cm}
\end{tabular}
\caption[]{
 Effect of the selection cuts on data, 
 simulated  background and simulated signal events.
 The two main background contributions are detailed.
 Efficiencies are given for a signal of \MH = 114~\GeVcc.
 Candidates selection indicates the number of events used as input 
 to the confidence level calculations. 
 The tight selection 
 is obtained after a further cut  in the corresponding discriminant
 variable, and corresponds to  
 the one used in the mass plot (Fig.~\ref{fig:mass_pl}).}
\label{ta:hzsum205}
\end{center}
\end{table}

\subsection{Leptonic channels}

 Higgs boson searches in events with jets and leptons
follow the analysis applied to the 1999 data~\cite{pap99}, 
which included a
\sqrts\ dependence in the corresponding preselections. 
 The effect of the selections on data and simulated samples  
is detailed in Table~\ref{ta:hzsum205}.
Good agreement between data and background simulation 
at the preselection level
is observed in all the leptonic channels.

In the \hee\ channel, 7 candidate events are selected in the
data, for a total expected background of 
\mbox{$11.6$} events coming mainly from the \eeqq\ process.
In the \hmm\ channel, 7 events are selected and   
\mbox{$10.6$} background events are expected coming mainly from the \mmqq\ process. 
Both channels use the
${\mathrm b}$-tagging value as the discriminant variable and 
the fitted hadronic mass in the two-dimensional 
calculation of the confidence levels.
One of the \hee\ and two of the \hmm\ candidates have 
a significant ${\mathrm b}$-tagging value 
but are kinematically compatible with the \ZZ~ hypothesis.

In the \ttqq\ channel, 5 candidates are selected,
while \mbox{$6.0$} are expected from the Standard Model  background, 
which is dominated by the \ZZ\ into \tautauqq\ process. Two events are selected after a cut on the 
discriminant likelihood at 0.1; 
neither has a high value for the rescaled mass.

\subsection{Higgs boson searches in four-jet events}
\label{sec:4jet}
  Higgs boson searches in fully hadronic final states start with a
common four-jet preselection~\cite{pap98,pap97}, which
eliminates hard radiative events and reduces the \qqg\
and $\Zz\gamma^*$ background, forcing all selected events 
into a four-jet topology with the DURHAM algorithm.

The performance of the DELPHI ${\mathrm b}$-tagging procedure 
in the four-jet analysis was specially optimized
and enhanced by taking into account the dependence on additional variables 
related to the kinematical properties of b-hadrons produced in decays of the
Higgs boson. These variables, defined for each jet in the event, are:
   the polar angle of the jet direction,
   the jet energy,
   the charged multiplicity of the jet,
   the angle to the nearest jet,
   the average transverse momentum of charged particles with respect
   to the jet direction,
   the number of particles with negative impact parameter and
  the invariant mass of the jet.
Including this dependence in the tagging algorithm significantly improves
the rejection of the light quark background.
The global ${\mathrm b}$-tagging value of the event is defined as
the maximum ${\mathrm b}$-tagging value for any di-jet in the event,
computed as the sum of the corresponding jet
${\mathrm b}$-tagging values.

The final discriminant variable used in the four-jet channel is 
defined as the output of an artificial neural 
network
(ANN) which combines 13 variables. 

The first variable is the global ${\mathrm b}$-tagging value of the event.

The next four  variables rely on kinematics and
test the compatibility of the event with the hypotheses of \WW\ and 
\ZZ\ production to either 4 or 5 jets. 
Constrained fits are used 
to derive the probability density function measuring the compatibility 
of the event kinematics with the production of two objects of any masses. 
This two-dimensional probability, the ideogram probability~\cite{ww183}, 
is then folded with the expected mass distributions for the \WW\ and \ZZ\
processes, respectively.

Finally, the last eight  input variables 
intended to reduce the \qqg\ contamination are 
the sum of the second and fourth Fox-Wolfram moments,
the product of the minimum jet energy and the minimum opening angle between
any two jets,
the maximum and minimum jet momenta,
the sum of the multiplicities of the two jets with lowest multiplicity,
the sum of the masses of the two jets with lowest masses,
the minimum di-jet mass and
the minimum sum of the cosines of the opening angles of the two di-jets
when considering all possible pairings of the jets.
In the previous analysis~\cite{pap99} these eight variables were 
separately combined in an anti-QCD artificial neural network.

\begin{figure}[htbp]
\begin{center}
\epsfig{figure=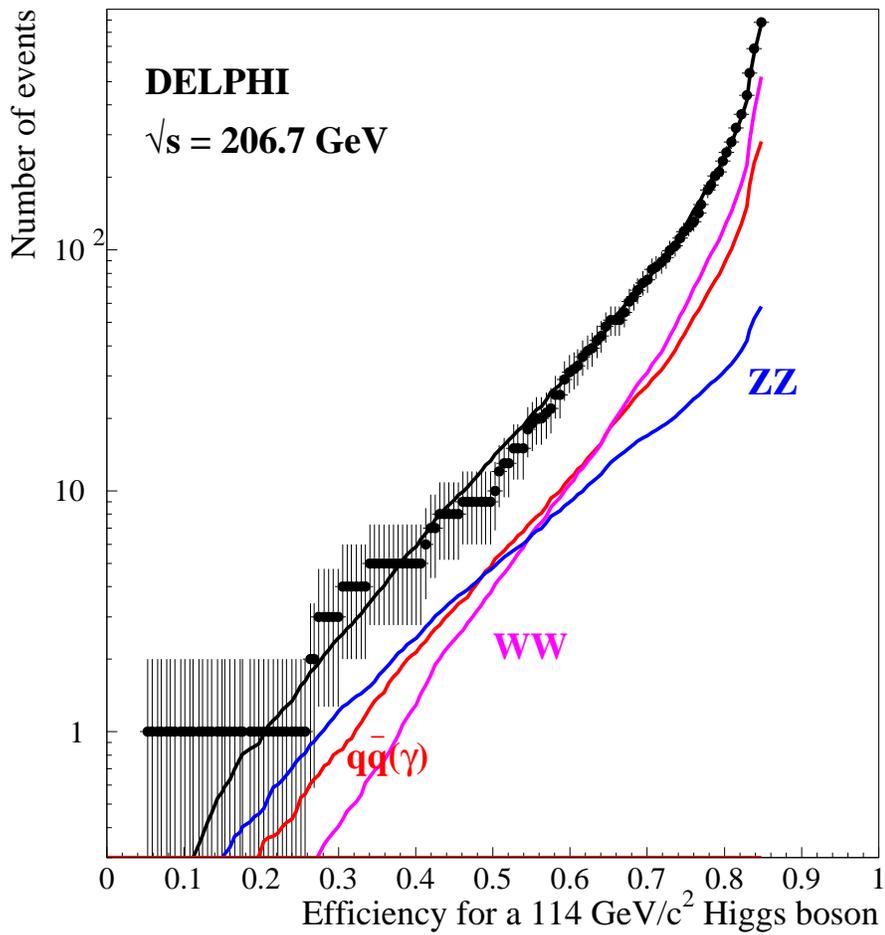,width=13cm} 
\caption{\hqq channel: 
  Expected Standard Model background rate at a centre-of-mass energy
  \rs~=206.7~\GeV\  as a 
  function of the efficiency for a 114~\GeVcc\ Higgs mass signal when
  varying the cut on the neural network variable. 
  The different background contributions are shown summed and separately. 
  Dots stand for data.} 
\label{fig:hqq_disc}
\end{center}
\end{figure}

Fig.~\ref{fig:hqq_disc} shows the 
performance  of the final discriminating variable
in the efficiency-background  plane
for a 114~\GeVcc\ signal
at \rs~=206.7~\GeV .


The choice of the Higgs di-jet makes use of both the kinematical 5C-fit 
probabilities and the ${\mathrm b}$-tagging information 
in the event~\cite{pap97}. 
The likelihood pairing function,

$  {\cal P}_b^{j_1} \cdot {\cal P}_b^{j_2} \cdot (
 (1-R_b^Z-R_c^Z) \cdot  {\cal P}_q^{j_3} \cdot {\cal P}_q^{j_4}
  +R_b^Z         \cdot  {\cal P}_b^{j_3} \cdot {\cal P}_b^{j_4}
  +R_c^Z         \cdot  {\cal P}_c^{j_3} \cdot {\cal P}_c^{j_4})
  \cdot P_{j_3,j_4}^{5C} $

\noindent is calculated for each of the six possibilities to combine the jets
$j_1$,$j_2$, $j_3$ and $j_4$.
${\cal P}_b^{j_i},{\cal P}_c^{j_i},{\cal P}_q^{j_i}$ are the 
probability densities of getting 
the observed ${\mathrm b}$-tagging value for the jet $j_i$ 
when originating from 
a $b$, $c$ or light quark, estimated from simulation.
$R_b^Z$ and $R_c^Z$ are the hadronic branching fractions
of the $Z^0$ into $b$ or $c$ quarks, and $P_{j_3,j_4}^{5C}$ is the probability 
corresponding to the kinematical 5C-fit with the jets $j_3$ and $j_4$ 
assigned to the $Z^0$. 
The pairing that maximises this function is selected.
The proportion of right matchings for the Higgs di-jet, 
estimated in simulated signal events with 114~\GeVcc\ mass, 
is around 53\% at preselection level, increasing to above 70\% at the 
tighter level, keeping a low rate of wrong pairings 
for \ZZ~ background events.

\begin{figure}[htbp]
\begin{center}
\epsfig{figure=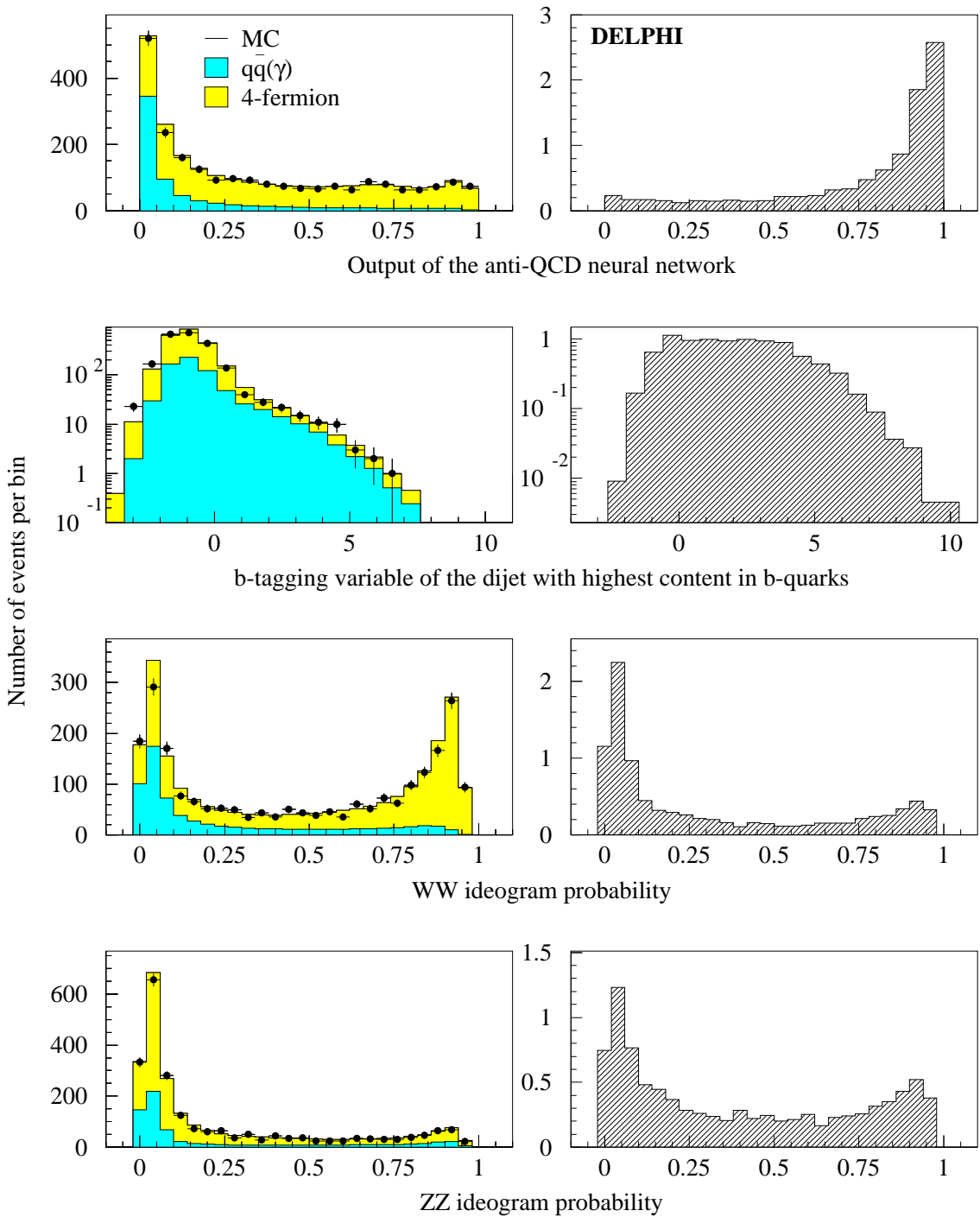,width=17cm}
\caption{ \hqq channel: 
  distributions of relevant analysis variables at the preselection level. 
  The eight variables used to reduce the \qqg\ background are 
  summarized by the output of the anti-QCD neural network.
  Data at \rs~=~200-209~\GeV\ (dots) are compared with Standard Model background 
  expectations (left-hand side histograms) and with the expected distribution
  for a 114~\GeVcc\ Higgs mass signal (right-hand side histogram).}
\label{fig:hqq_1}
\end{center}
\end{figure}

\begin{figure}[htbp]
\begin{center}
\epsfig{figure=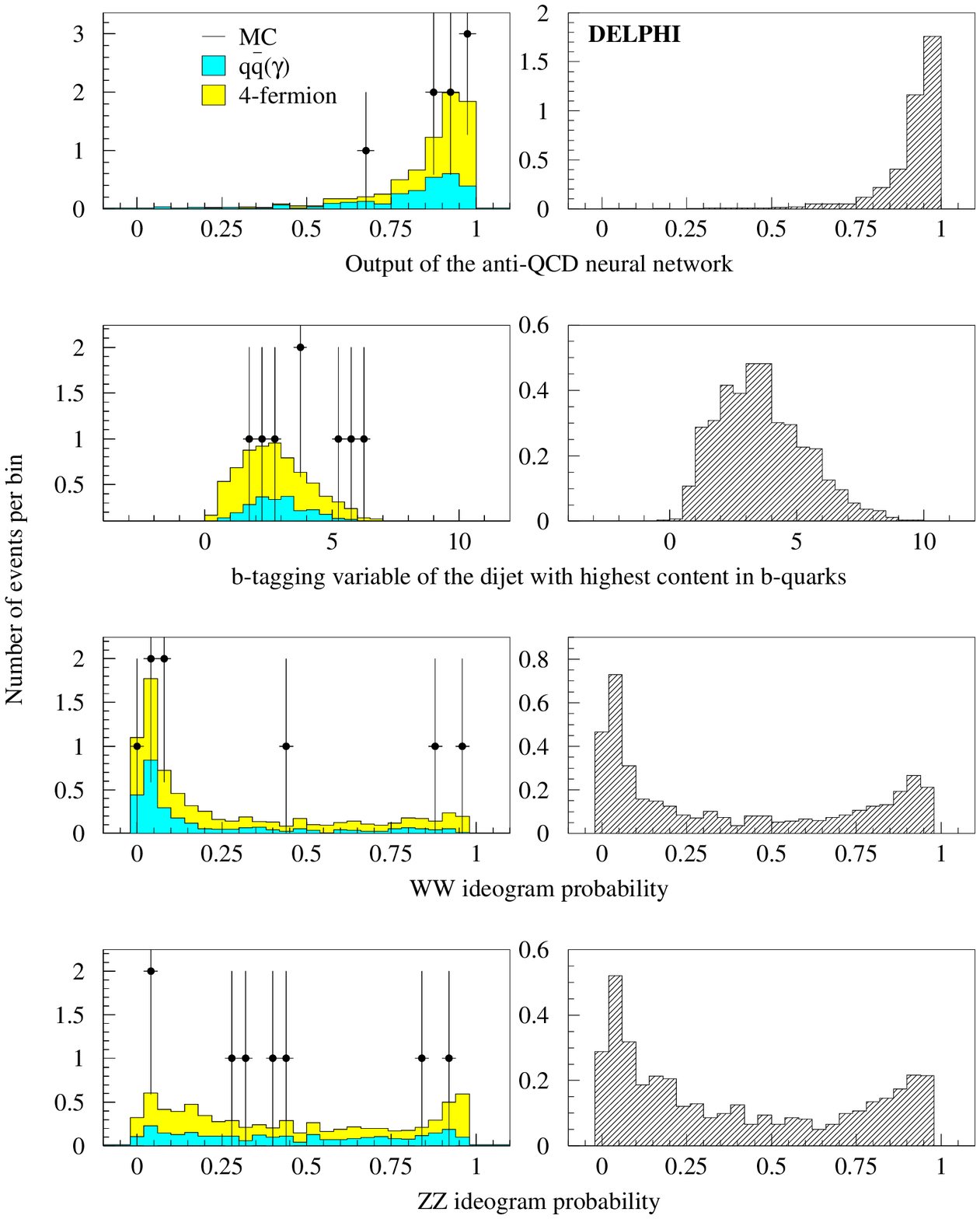,width=17cm}
\caption{ \hqq channel: 
same distribution as in Fig.\ref{fig:hqq_1} but at the tight selection level.}
\label{fig:hqq_2}
\end{center}
\end{figure}

The good agreement between data and 
background simulation after the four-jet
preselection is illustrated 
in  Fig.~\ref{fig:hqq_1} which shows the distributions 
of the global ${\mathrm b}$-tagging, the two ideogram probabilities
for the configuration with 4 jets, and the output of
the anti-QCD ANN. 
The results for the different selection levels 
are given in Table~\ref{ta:hzsum205}.
The tighter cut at an ANN value of $0.7$ selects $8$ events in data
while $7.4$ are expected from the background simulation.
Fig.~\ref{fig:hqq_2} shows the previous variables at this level.

\subsection{Confidence level estimation}

The confidence levels for the background (CL$_{\rm b}$) and signal
plus background (CL$_{\rm s+b}$) hypotheses are defined as the probability
in the two cases of observing a likelihood ratio \likear,
greater than or equal to that measured in the data~\cite{alex}. 
The confidence level for the
signal case is calculated consistently with the LEP Higgs Working group
using the conservative ratio CL$_{\rm s}$ = CL$_{\rm s+b}$/CL$_{\rm b}$.

The likelihood ratio for a given Higgs mass hypothesis is defined as 
$\ln(\likear) = -S + \Sigma_i \ln(1+s_i/b_i)$ 
where $S$ is the total expected signal, and $s_i$ and $b_i$ are the 
signal and background probability densities for each candidate $i$, 
calculated using two-dimensional information,
where one dimension is the reconstructed Higgs boson mass and the other is
the channel dependent discriminant variable.

These densities are represented as two-dimensional histograms which
are derived from the simulation samples described in section 1.1. These
distributions are
then smoothed using a two-dimensional kernel, which is essentially Gaussian
but with a small longer tailed component. The width of the kernel varies
from point to
point, such that the statistical uncertainty on the estimated background
is never more than 30\%. The same width is applied to background
and all signal samples to eliminate the possibility of the smearing itself
increasing the estimated signal to background ratio. 
Finally the distribution is
reweighted so that when projected onto either axis it has the same
distribution 
as would have been observed if the smoothing had been only in one dimension.
This makes better use of the simulation statistics if there are features which
are essentially one dimensional, such as mass peaks, and it has been verified
that the systematic errors introduced are significantly smaller
than the statistical ones.


The resultant two-dimensional distributions
are then linearly interpolated from the simulation conditions 
to the appropriate
beam energy and Higgs mass hypotheses.

\section{Results}
\label{sec:smresults}

\begin{figure}[htbp]
\begin{center}
\epsfig{figure=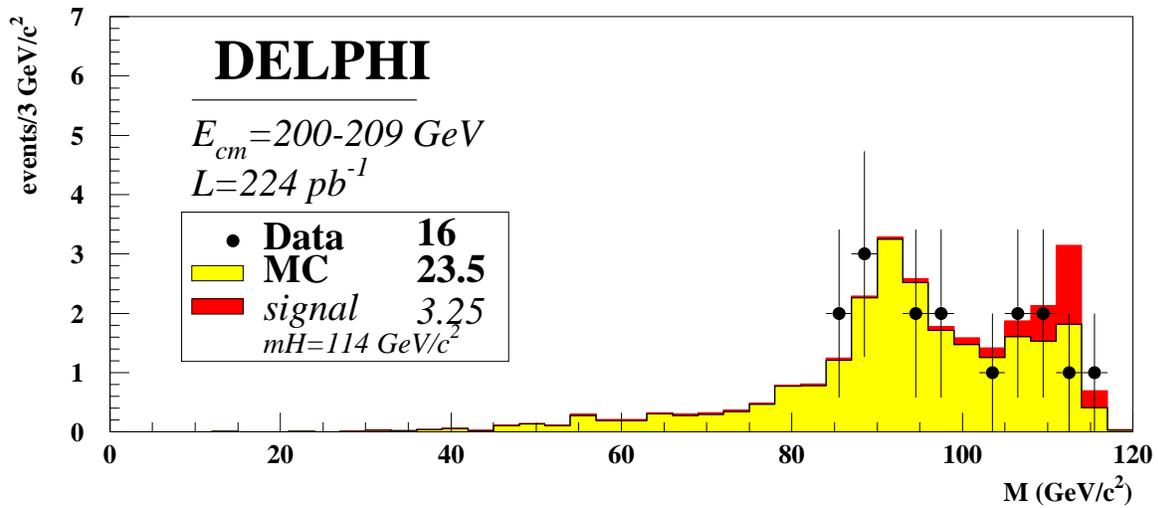,width=17.cm}
\vspace{-5cm.}
\caption[]{
Distribution of the reconstructed mass of the candidates
when combining all \ZH\ analyses 
at 200-209~\GeV\ in the year 2000. 
Data (dots) are compared with the Standard Model background expectations 
(light shaded histogram)
and with the normalised 114~\GeVcc\ signal spectrum added to the background 
contributions (dark shaded histogram).}
\label{fig:mass_pl}
\end{center}
\end{figure}

The distribution of the reconstructed Higgs boson mass summed over all
channels, at the level
of the tight selection,
is presented in Fig.~\ref{fig:mass_pl}. 

The limit on the Standard Model Higgs boson mass is set combining
the data analysed in the previous sections with those taken at 
lower energies, namely 
161 and 172~\GeV~\cite{pap96}, 183~\GeV~\cite{pap97},
189~\GeV~\cite{pap98} and 192-202~\GeV~\cite{pap99}.

\begin{figure}[htbp]
\begin{center}
\epsfig{figure=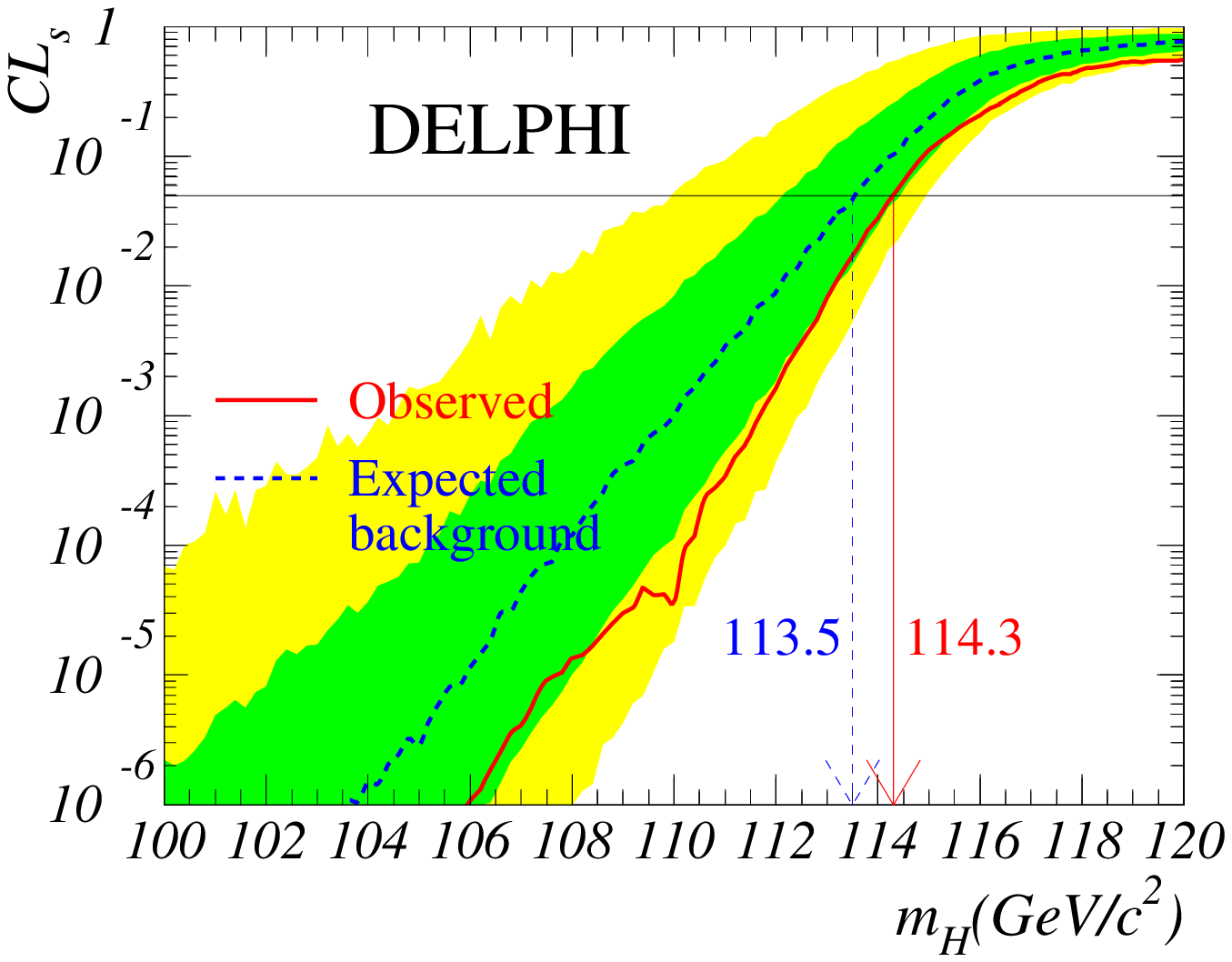,width=13.cm}
\epsfig{figure=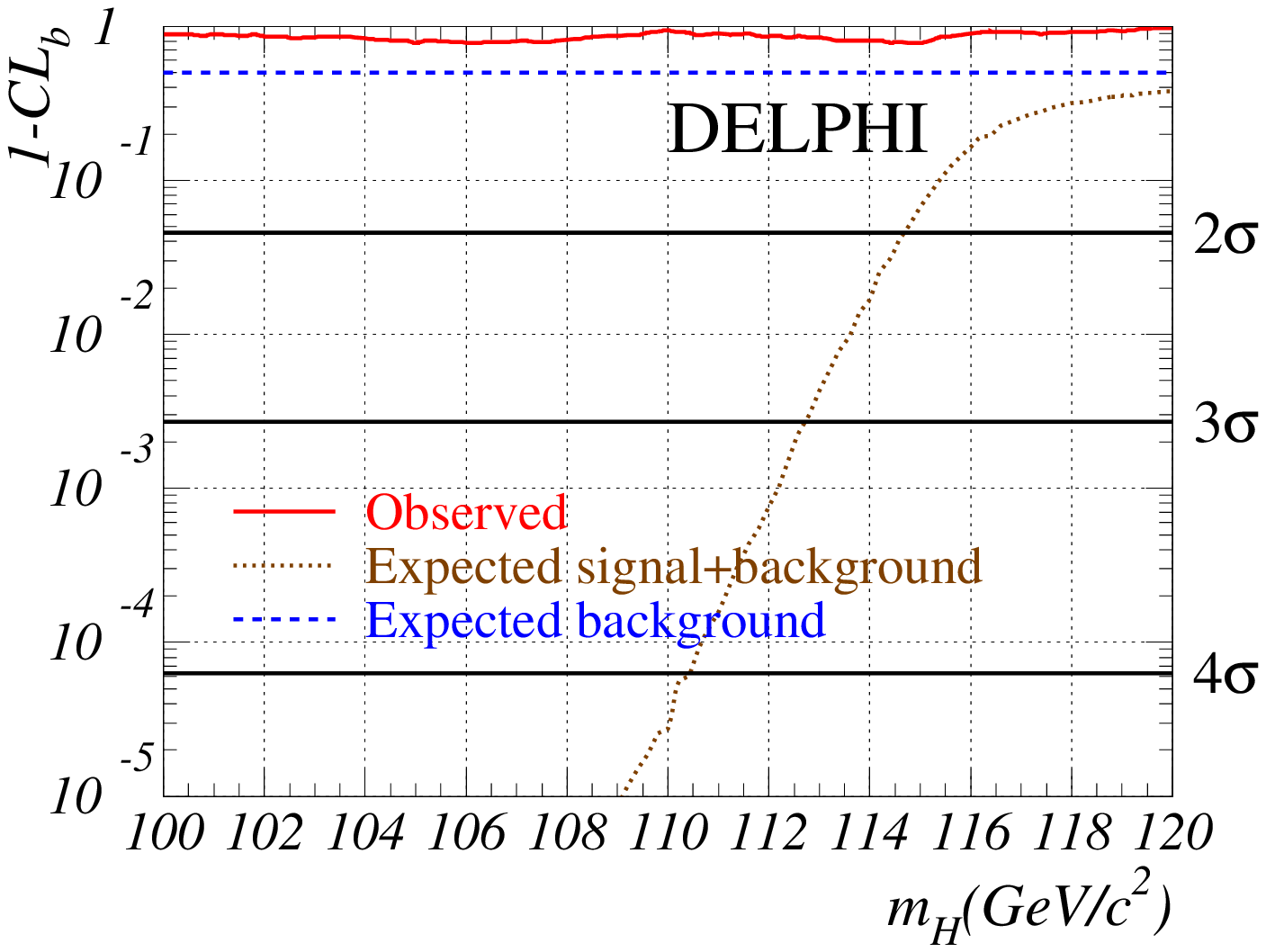,width=13.cm}
\caption[]{
Confidence levels as a function of \MH.
Curves are the observed (solid) and expected median 
(dashed) confidences from background-only experiments
while the bands correspond to the 68\% and 95\% confidence intervals 
from background-only experiments.
Top: CL$_{\rm s}$, the
confidence level for the signal hypothesis 
 as a function of \MH.
The intersections of the 
curves with the horizontal line at 5\% define the 
observed and expected 95\% CL lower limits 
on \MH at 114.3 and 113.5~\GeVcc\, respectively.
Bottom: 1-CL$_{\rm b}$ for the
background hypothesis.
Also shown here
is the curve of the median confidence as expected for a signal of mass given
in abscissa (dotted line).
}
\label{fig:cls_sm}
\end{center}
\end{figure}

\begin{figure}[htbp]
\begin{center}
\epsfig{figure=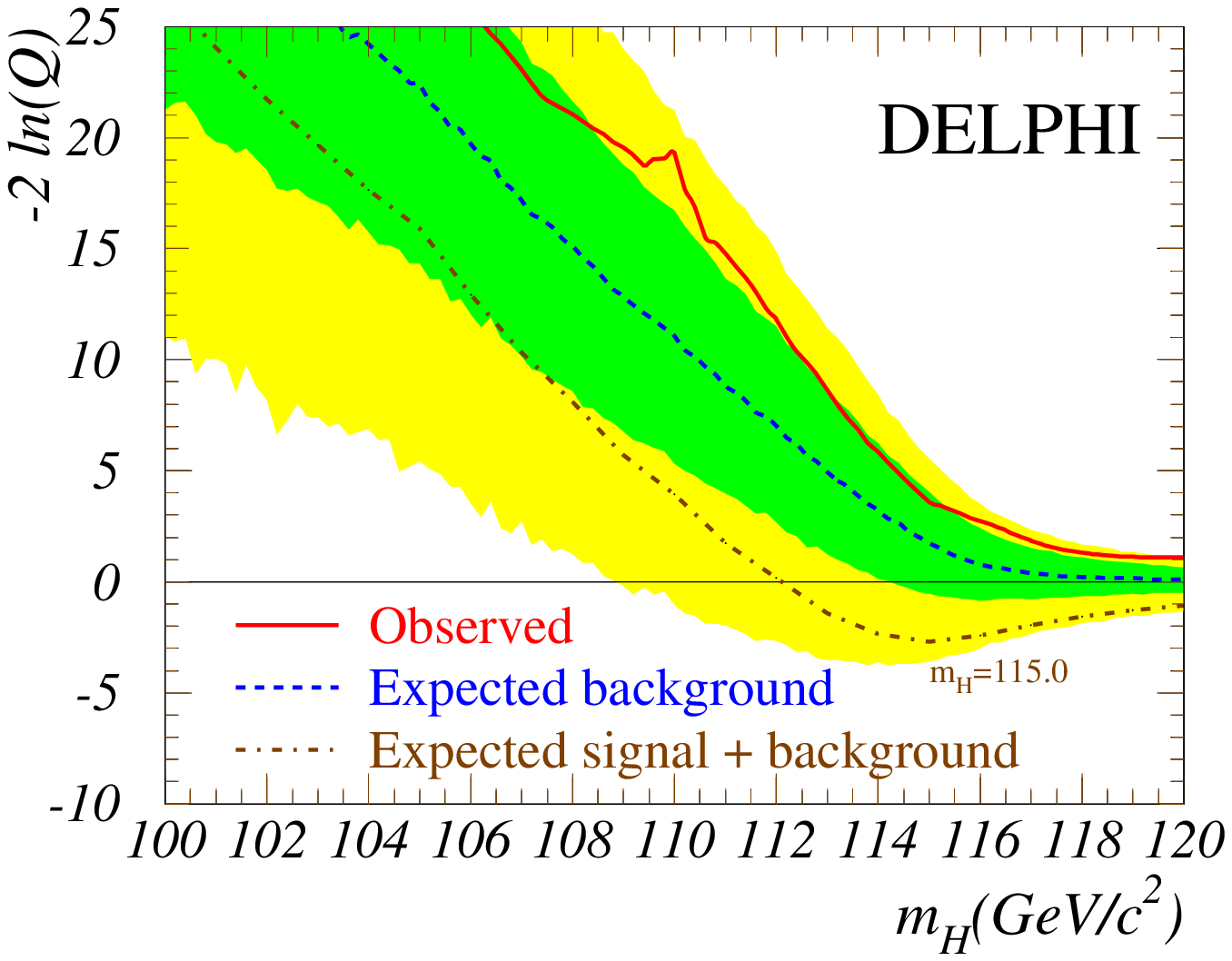,width=13.cm}
\vspace{-0.5cm.}
\epsfig{figure=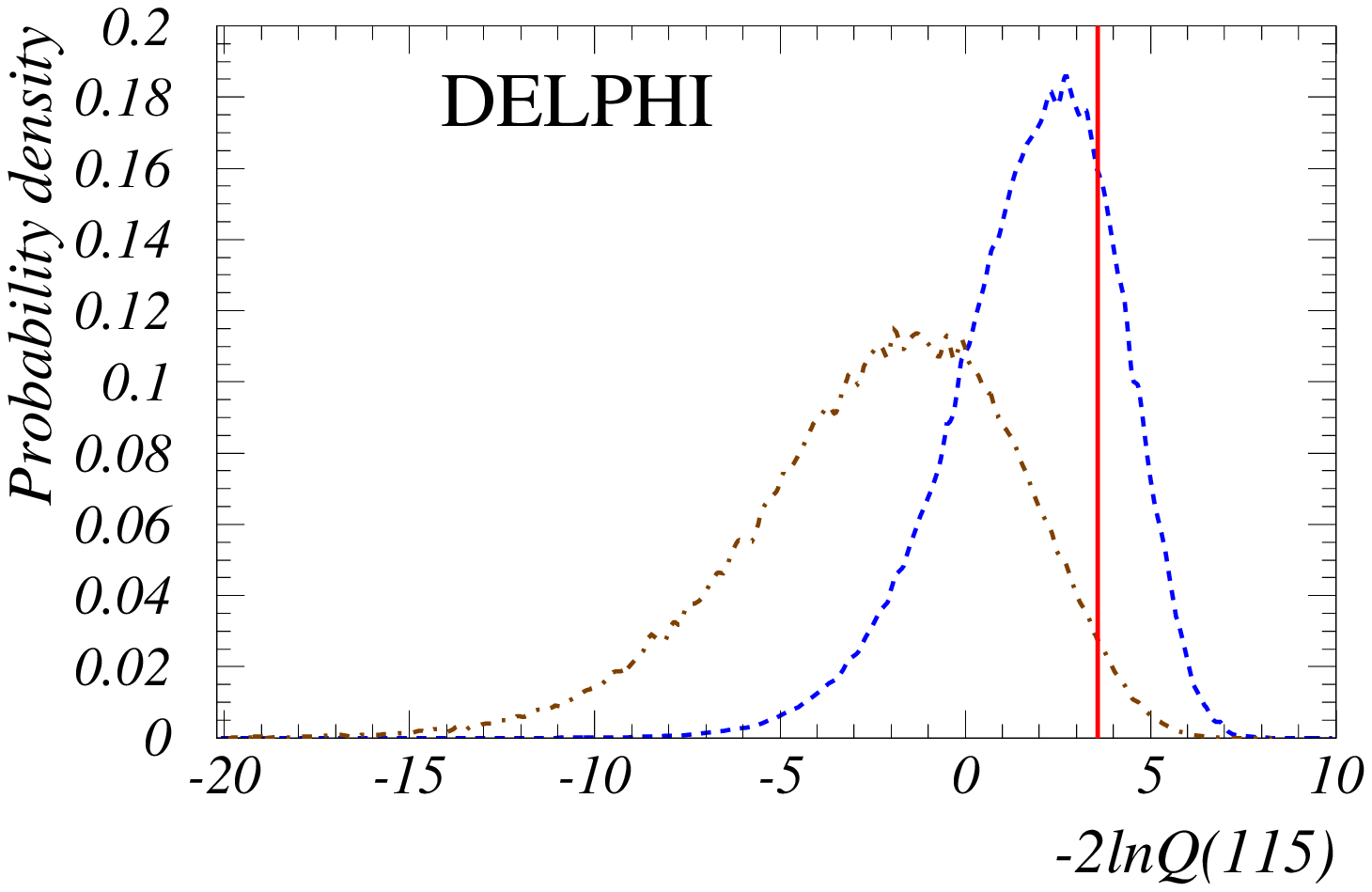,width=13.cm}
\caption[]{
Top: the test-statistic (negative log-likelihood ratio)
as a function of \MH.
The observed value, full line, is compared to 
the expectation for the background only hypothesis, 
represented by the dashed line and the symmetric 68\% and 95\% 
probability shaded bands. 
The dot-dashed line shows the average expected 
result for a hypothetical Higgs mass of 115~\GeVcc.
Bottom: vertical slice of the previous plot for a
mass value of 115~\GeVcc, showing the sensitivity of the DELPHI result
to this hypothesis.
The dot-dashed line shows the expected
distribution for signal plus background, the dashed line that for background
only. The vertical line represents the data. The fractional area below the
dashed curve and to the right of the data is CL$_{\rm b}$;
for the dot-dashed curve it is CL$_{\rm (s+b)}$.
}
\label{fig:xis_sm}
\end{center}
\end{figure}

The confidence level for the signal hypothesis CL$_{\rm s}$
is shown in Fig.~\ref{fig:cls_sm},
as well as the confidence level for the background hypothesis 
in the form 1-CL$_{\rm b}$.
A slight deficit with respect to the expected background is
observed, and a 95\% CL lower limit on the mass is set at 
114.3~\GeVcc\, while  the expected median is 113.5~\GeVcc\ .
The test-statistic (negative log-likelihood ratio)  
is shown in Fig.~\ref{fig:xis_sm}.

It has been noticed that the combined LEP result~\cite{LEPHWG_2000}
is better described if a Higgs boson with mass 115~\GeVcc\ is
present. 
For such a signal, the DELPHI CL$_{\rm s+b}$ value is 3\%, 
while the CL$_{\rm b}$ is 23\%. 
The CL$_{\rm s}$ for this hypothesis is 12\%,
so that the present data are not incompatible with 
the existence of a Higgs boson with this mass.
This can also be seen in Fig.~\ref{fig:xis_sm}, where 
the result is compared with the probability density for
background and  background plus signal experiments.
 
\section{Conclusions}

The data taken by DELPHI at 200-209~\GeV\ in the year 2000 have been analysed 
to search for the Standard Model Higgs boson. 
The data for all channels is compatible with
 expectations from the Standard Model background.
In combination with previous DELPHI results at lower centre-of-mass energies, 
a lower limit at 95\% CL on the mass of the 
Standard Model Higgs boson is set at \mbox{114.3 ~\GeVcc}, 
while the expected median limit is \mbox{113.5 ~\GeVcc}.

\subsection*{Acknowledgements}
\vskip 3 mm
We are extremely grateful to the members of CERN-SL Division for
their expertise and dedication that has allowed the LEP energy to be increased
well beyond the design value, with a consequent extension of the search for
the Higgs boson.
 We are greatly indebted to our technical 
collaborators
 and to the funding agencies for their
support in building and operating the DELPHI detector.\\
We acknowledge in particular the support of \\
Austrian Federal Ministry of Science and Traffics, GZ 616.364/2-III/2a/98, \\
FNRS--FWO, Flanders Institute to encourage scientific and technological 
research in the industry (IWT), Belgium,  \\
FINEP, CNPq, CAPES, FUJB and FAPERJ, Brazil, \\
Czech Ministry of Industry and Trade, GA CR 202/96/0450 and GA AVCR A1010521,\\
Commission of the European Communities (DG XII), \\
Direction des Sciences de la Mati$\grave{\mbox{\rm e}}$re, CEA, France, \\
Bundesministerium f$\ddot{\mbox{\rm u}}$r Bildung, Wissenschaft, Forschung 
und Technologie, Germany,\\
General Secretariat for Research and Technology, Greece, \\
National Science Foundation (NWO) and Foundation for Research on Matter (FOM),
The Netherlands, \\
Norwegian Research Council,  \\
State Committee for Scientific Research, Poland, 2P03B06015, 2P03B11116 and
SPUB/P03/DZ3/99, \\
JNICT--Junta Nacional de Investiga\c{c}\~{a}o Cient\'{\i}fica 
e Tecnol$\acute{\mbox{\rm o}}$gica, Portugal, \\
Vedecka grantova agentura MS SR, Slovakia, Nr. 95/5195/134, \\
Ministry of Science and Technology of the Republic of Slovenia, \\
CICYT, Spain, AEN99-0950, AEN99-0761  \\
The Swedish Natural Science Research Council,      \\
Particle Physics and Astronomy Research Council, UK, \\
Department of Energy, USA, DE--FG02--94ER40817, \\




\clearpage

\newpage


\end{document}